\documentclass[aps,twocolumn,pre,superscriptaddress,showpacs,showkeys,floatfix,nofootinbib]{revtex4-1}
\usepackage{graphicx,amssymb,amsmath,bm,multirow}
\usepackage[usenames,dvipsnames,svgnames,table]{xcolor}
\usepackage{grffile}
\newcommand{\boolvar}{b}\newcommand{\sigmapert}{\sigma}
\newcommand{\sigmaI}{{\sigma^{(I)}}}
\newcommand{\sigmaE}{{\sigma^{(E)}}}
\newcommand{\REV}[1]{{ \color{black} #1} }

\begin{document}
\title{Evolution of regulatory networks towards adaptability and stability in a changing environment}
\author{Deok-Sun \surname{Lee}}
\email{deoksun.lee@inha.ac.kr}
\affiliation{Department of  Physics, Inha University, Incheon 402-751, Korea}
\date{\today}

\begin{abstract}
Diverse biological networks exhibit universal features distinguished from those of random networks, 
calling much attention to their origins and implications. 
Here we propose a minimal evolution model of Boolean regulatory networks, 
which evolve by selectively rewiring links  towards enhancing adaptability to a changing environment 
and stability against dynamical perturbations. 
We find that sparse and heterogeneous connectivity patterns emerge, which show qualitative  
agreement  with real transcriptional regulatory networks and metabolic networks. 
The characteristic scaling behavior of stability reflects the balance between robustness and flexibility. 
The scaling of fluctuation in the perturbation spread  shows a dynamic crossover, 
which is analyzed by investigating separately the stochasticity of internal dynamics 
and the network structures different depending on the evolution pathways.  
Our study delineates how the ambivalent pressure of evolution shapes biological networks, 
which can be helpful for studying general complex systems interacting with environments. 
\end{abstract}
\maketitle

\section{Introduction}

The global organization of complex molecular interactions 
within  and across cells is being disclosed 
by the graph-theoretic approaches~\cite{Barabasi:2004uq,tilee02,Rual:2005kx,Ma22012003}.  
The obtained cellular networks exhibit  universal topological features which are rarely found in random 
networks, such as broad degree distributions~\cite{Jeong:2000kx} and 
high modularity~\cite{shenorr02}. 
Their origins and implications to cellular and larger-scale functions have thus been of great interest. 
Diverse network models based on simple mechanisms of adding 
and removing nodes and links have been proposed ~\cite{barabasi99,Vazquez:2003vn,Yamada:2009qe}. 
Those models capture the common aspects, like the 
preferential attachment~\cite{Barabasi:1999ys}, of biological processes 
such as  the duplication, divergence, and recruitment of genes, proteins, and enzymes,  
and successfully reproduce the empirical features of biological networks, suggesting that 
the former can be the origin of the latter.  
Yet it remains to be explored what drives such construction and remodeling of 
biological networks functioning in living organisms. 
A population of living organisms  find the typical  
architecture and function of their cellular networks  
changing with time.   
Such changes on long time scales are made by 
the organisms of different traits, giving birth to their descendants with different chances, 
that is, by evolution~\cite{fisher1999genetical,Orr:2005fk}.   
Therefore, it is desirable to investigate how the generic  features of 
evolution lead to the emergence of the common features of biological networks.

Living organisms are required to possess adaptability and stability simultaneously~\cite{wagnerbook}. 
To survive and give birth to descendants in fluctuating environments,  
the ability to adjust to a changed environment is essential~\cite{beaumont09}, 
which leads to, e.g., phenotypic diversity and 
the advantage of bet-hedging strategy~\cite{beaumont09}. 
At the same time, the ability to maintain the constant structure and perform 
routine important functions regularly, 
such as cell division and heat beats, is highly demanded. 
Therefore, in a given population, the cellular networks supporting higher adaptability and stability 
are more likely to be inherited,  which  
leads the representative topology and function of the cellular networks 
to evolve over generations.     

Here we study how
such evolutionary pressure shapes the biological networks. 
We propose a network model, in which  links are rewired such that both 
adaptability and stability are enhanced. The dynamics of the network is 
simply represented by the Boolean variables assigned to each node 
regulating one another~\cite{kauffman69}. 
The Boolean networks have been instrumental for studying 
the gene transcriptional regulatory networks~\cite{babu04} and the metabolic networks~\cite{Ghim2005401}. 
This model network is supposed to represent the network structure typical of 
a population.
The evolution of Boolean networks towards enhancing 
adaptability~\cite{kauffman86, stern99,oikonomou06,stauffer09,Greenbury201048},
stability~\cite{Wagner19961008,bornholdt98,szejka07,Sevim2008323,1367-2630-11-3-033005,Esmaeili2009127,mihaljev09,PhysRevE.81.021908,peixoto12}, 
or both~\cite{10.1371/journal.pcbi.1002669} has been studied, 
mostly by applying the genetic algorithm or similar ones to a group of small networks. 
In particular, the model networks which evolve by rewiring links towards 
 local dynamics being neither active nor inactive 
have been shown to 
reproduce the critical global connectivity and many of the universal features of real-world biological 
networks~\cite{bornholdt00, rohlf07, rohlf08, liu06}, demonstrating the close relation 
between evolution and the structure of biological networks.
However, the evolutionary evaluation and selection are made for each whole organism, not for part of it.  
In the simulated evolution of our model, the adaptability and the stability of 
the  {\it global} dynamical state are evaluated in the wild-type network 
and its mutant network, differing by a single link from each other, and 
the winner of the two becomes the wild-type in the next step. 
The study of this model leads us to find 
that sparse and heterogeneous connectivity patterns emerge, which are
consistent  with the gene transcriptional regulatory networks and the metabolic networks of diverse species. 
The scaling behavior of stability with respect to system size suggests that the evolved networks are critical, 
 lying at the boundary between the inflexible ordered phase and the unstable chaotic phase. 

Our study also shows how the nature of fluctuations and correlations changes by evolution. 
The extent of perturbation spread characterizing the system's stability 
fluctuates over different realizations of evolution. 
The fluctuation turns out to  scale linearly with the mean in the stationary state of evolution 
while the square-root scaling holds in the transient period.  
We argue that this dynamic crossover is rooted in the variation of the combinatorial impacts 
of the structural fluctuation, driven by evolution, and the  internal stochasticity.  
The scaling of the correlation volume, representing 
the typical number of nodes correlated with a node, is another feature of the evolved networks.
Our results thus show the universal impacts of biological evolution on the 
structure and function of biological networks 
and illuminate the nature of correlations and fluctuations in such evolving systems 
distinguished from randomly-constructed or other artificial systems.

The paper is organized as follows. 
The network evolution model is described in detail in Sec.~\ref{sec:model}. 
The emergent structural and functional features 
are presented in Sec.~\ref{sec:evolution}. 
In Sec.~\ref{sec:generalized},  we represent  the Hamiltonian approach to a generalized model, 
including our model in a limit, 
and show the robustness of the obtained results. 
The scaling behaviors of the fluctuation of  perturbation spread and the correlation volume are 
analyzed in Secs.~\ref{sec:fluctuation} and \ref{sec:correlation}, respectively. 
We summarize and discuss the results of our study in Sec.~\ref{sec:discussion}.

\section{Model}
\label{sec:model}
 
We consider a network in which the node activities are regulated by one another.
The network may represent the transcriptional regulatory network of genes, 
in which the transcription of a gene is affected by the  
transcriptional factors encoded from other genes, 
or the metabolic network of metabolites and reactions, 
the concentrations and fluxes of which are correlated. 
Various cellular functions are based on 
those elementary regulations.   
The model network does not mean that of  a specific organism 
but is representative of the cellular networks 
of a population of organisms, which evolve with time.  
\REV{In our model the network evolution is made by adding or removing links, 
representing the establishment of new regulatory inputs or the loss of existing 
targets possibly caused by point 
mutations in the regulatory or coding regions of DNA~\cite{10.1371/journal.pcbi.1002669, babu04}.}

To be specific, we consider a network $G$ of $N$ nodes which are assigned Boolean variables 
$\boolvar_{i}=\pm 1$ for $i=1,2,\ldots,N$.
$b_i$ represents whether a  node $i$ is active or inactive in terms of the transcription of 
the messenger RNA, the flux of the corresponding chemical reaction, or the concentration of 
the  metabolite. 
The global dynamical state  is represented by  $\Sigma = \{ b_1, b_2, \ldots, b_N\}$. 
Initially $L_0$ directed links are randomly wired and $b_i$'s are  set to $1$ or $-1$ randomly. 
A link from node $j$ to node $i$, with the adjacency matrix $A_{ij}=1$, indicates 
the regulation of the activity of $i$ by $j$~\cite{babu04,Ghim2005401}.  
$b_{i}(\tau+1)$ of node $i$ at  the microscopic time step $\tau+1$ is determined by its regulators at $\tau$ as   
\begin{equation}
\boolvar_{i}(\tau+1) = F_{i}(\{ \boolvar_{j}(\tau)|A_{ij}=1\}),
\label{eq:boolean_evol}
\end{equation}
where $F_i$ is  the time-constant regulation function for node $i$, taking a value 
$1$ or $-1$ for each of all the $2^{k_i}$ states of $k_i$ regulators with $k_i = \sum_{j} A_{ij}$.
A  target state 
$\Sigma^{\rm (target)} = \{ \boolvar_1^{\rm (target)}, \boolvar_2^{\rm (target)}, \ldots, \boolvar_N^{\rm (target)}\}$ 
is demanded of the network by the environment and the distance between $\Sigma$ and $\Sigma^{\rm (target)}$ 
quantifies the adaptation to the environment.

\begin{figure}
\begin{center}
\includegraphics[width=0.7\columnwidth]{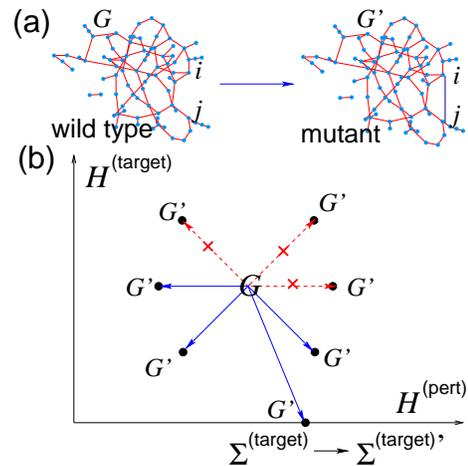}
\caption{(Color online) Evolving network model. (a) A mutant $G'$ is generated by adding or removing a link 
randomly in the wild-type $G$, here between nodes $i$ and $j$. 
(b) The transition from $G$ to $G'$ happens if  $H^{\rm (target)}_{G'}<H^{\rm (target)}_{G}$  
or if  $H^{\rm (pert)}_{G'}\leq H^{\rm (pert)}_{G}$ and $H^{\rm (target)}_{G'}=H^{\rm (target)}_{G}$. 
A new target state ${\Sigma^{\rm (target)}}'$ is generated if $H^{\rm (target)}_{G',t}=0$.}
\label{fig:model}
\end{center}
\end{figure}

The dynamical state $\Sigma(\tau)=\{b_1(\tau), b_2(\tau),\ldots, b_N(\tau)\}$ is updated 
every microscopic time step $\tau$ as in Eq.~(\ref{eq:boolean_evol}). 
Also the  structure of the network $G$, including its adjacency matrix $A$ 
and the regulating functions $\{F\}$, evolves on a longer time scale as follows. 
At $\tau = t\tau_m$ with $t=0,1,2,\ldots$ the macroscopic time step and $\tau_m$ a time constant, 
a  mutant network  $G'$ is generated, which is identical to the wild-type $G$ except that 
it has  one more or less link with  a different regulation function (See Fig.~\ref{fig:model}).
Then we let the dynamical state $\Sigma(\tau)$ evolve on $G$ and $G'$, respectively, 
for $t\tau_m\leq \tau < (t+1) \tau_m$. 
Due to their structural difference, the $\Sigma(\tau)$'s may evolve differently 
although they are set equal initially at $\tau=t\tau_m$. 
At $\tau = (t+1)\tau_m$, 
the adaptability and the stability of the time trajectories $\{\Sigma(\tau)|t\tau_m \leq \tau<(t+1)\tau_m\}$ 
on $G$ and $G'$ are evaluated  in terms of the Hamming distances, 
$H^{\rm (target)}_{G,t}, H^{\rm (target)}_{G',t}, H^{\rm (pert)}_{G,t}$, and $H^{\rm (pert)}_{G',t}$, 
where the first two characterize the adaptation to the environment 
and the latter two represent the typical extent of perturbation spread.
The winner of $G$ and $G'$ is determined in the way detailed below, 
which then becomes the wild-type $G$ 
for $(t+1)\tau_m \leq \tau< (t+2) \tau_m$ competing with  its mutant. These 
procedures are repeated for $t=0,1,2,\ldots$. 

The adaptability of a Boolean network $G$ at time $t$ is here 
quantified by the average Hamming distance between  $\Sigma(\tau)$ and 
a given target state  $\Sigma^{\rm (target)}$~\cite{kauffman86, stern99,oikonomou06,stauffer09,Greenbury201048} 
over a microscopic time interval as 
\begin{align}
H_{G,t}^{\rm (target)} = {1\over \tau_m - \tau_s}& \sum_{\tau = t\tau_m+\tau_s }^{(t+1)\tau_m}  H(\Sigma(\tau), \Sigma^{\rm (target)}), \nonumber\\
H(\Sigma, \Sigma^{\rm (target)}) &= {1\over N} \sum_{i=1}^N \left(1-\delta_{b_i, b_i^{\rm (target)}}\right)
\label{eq:Htarget}
\end{align}
where $\delta_{a,b}$ is the Kronecker delta function. $\tau_s$ is a microscopic-time constant 
such that the  Hamming distance $H(\Sigma(\tau),\Sigma^{\rm (target)})$  
is stationary for $t\tau_m +\tau_s \leq \tau < t\tau_m +\tau_m$.  
Another constant $\tau_m$ is set  to $\tau_m = 2\tau_s$,  
which is found to range from $38$ to $162$ for   $30\leq N\leq 800$ in our simulations. 
\REV{If smaller values of $\tau_m$ and $\tau_s$ were used, 
$H_{G,t}^{\rm (target)}$ in Eq.~(\ref{eq:Htarget}) would not represent the adaptability 
of the network  in the stationary state of the Boolean dynamics.}
The smaller $H_{G,t}^{\rm (target)}$ is, the closer the dynamical state on $G$ is likely to 
approach the target state, 
implying that  $G$ is more adaptable to a given environment. 
We compute $H_{G',t}^{\rm (target)}$ in the same way as in Eq.~(\ref{eq:Htarget}).

The stability in performing routine processes is another key requirement of life. 
Given that local perturbations can spread globally, 
the ability to suppress such perturbation spread can be a measure of stability~\cite{Wagner19961008,bornholdt98,szejka07,Sevim2008323,1367-2630-11-3-033005,Esmaeili2009127,mihaljev09,PhysRevE.81.021908,peixoto12}.
To quantify the stability of $G$ at time $t$, 
the difference between the original state $\Sigma(\tau)$ and the perturbed state 
$\Sigma^{\rm (pert)}(\tau)=\{b_1^{\rm (pert)}(\tau), b_2^{\rm (pert)}(\tau),\ldots, b_N^{\rm (pert)}(\tau)\}$ is measured. 
The  perturbed state  is obtained by flipping the states of $N/2$ 
randomly-selected $\boolvar$'s in $\Sigma(\tau)$ 
at $\tau=t\tau_m$ and then letting it evolve on $G$ for $t\tau_m \leq \tau< (t+1)\tau_m$.
Then we count the number of perturbed nodes, having 
$b_i\ne b_i^{\rm (pert)}$, as
 \begin{equation}
H_{G,t}^{\rm (pert)} = {1\over \tau_m - \tau_s} \sum_{\tau = t\tau_m+\tau_s }^{(t+1)\tau_m}  H(\Sigma(\tau), \Sigma^{\rm (pert)} (\tau)) 
\label{eq:Hpert}
\end{equation}
with the Hamming distance $H(\Sigma, \Sigma^{\rm (pert)})$  defined in Eq.~(\ref{eq:Htarget}).
$H^{\rm (pert)}_{G,t}$ represents the typical fraction of perturbed nodes; 
the smaller $H^{\rm (pert)}_{G,t}$ is,  the more stable the network $G$ is against 
dynamical perturbations. 
The stability of the mutant $G'$ is also computed in the same way.
We remark that the number of 
initial flipped variables can be changed  over a significant range without changing 
the main results.

The mutant $G'$ becomes the winner  (i) if $H_{G',t}^{\rm (target)}<H_{G,t}^{\rm (target)}$ ($G'$ is 
more adaptable than $G$)  or (ii) if $H_{G',t}^{\rm (pert)}< H_{G,t}^{\rm (pert)}$ ($G'$ is more stable 
than $G$) and $H_{G',t}^{\rm (target)}=H_{G,t}^{\rm (target)}$.  
If $H_{G',t}^{\rm (target)}=H_{G,t}^{\rm (target)}$ and $H_{G',t}^{\rm (pert)}= H_{G,t}^{\rm (pert)}$, 
the winner is chosen at random. 
Examples of the transition from $G$ to $G'$ are depicted  in Fig.~\ref{fig:model}. 
Finally, to model the changes of the environment, a new target state 
${\Sigma^{\rm (target)}}^{'}$ is generated if $H^{\rm (target)}$ of the winner is zero.  
Therefore our network evolution model 
represents the co-evolution of the structure and dynamics of the Boolean network on 
different time scales in a changing environment.

\section{Emergent features in structure and function}
\label{sec:evolution}

\begin{figure}
\begin{center}
\includegraphics[width=\columnwidth]{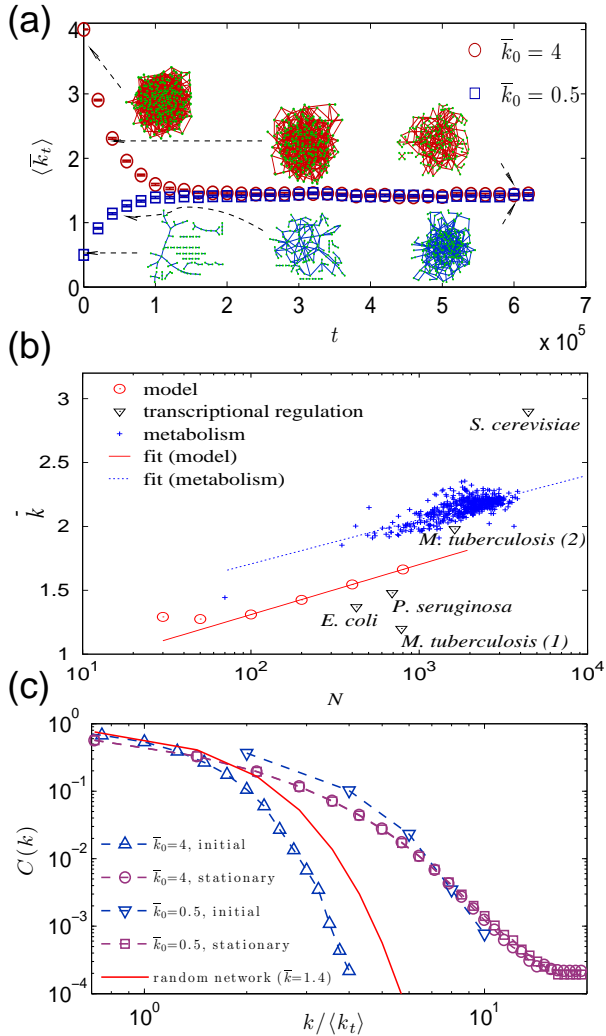}
\caption{(Color online) Emergence of a sparse and heterogeneous connectivity pattern. 
In simulations,  the initial number of links $L_0$ is set to 
$4N$ or $N/2$ giving  $\overline{k}_0=L_0/N = 4$ or $0.5$.  
For each $N$ and $L_0$, we run $\mathcal{N}$ independent simulations, each for $0\leq t\leq T$, 
where $T$ ranges from $4\times 10^4$ to $5\times 10^6$ and $\mathcal{N}=1000$ for $N\leq 200$, 
$\mathcal{N}=760$ for $N=400$, and $\mathcal{N}=22$ for $N=800$. 
$\langle \cdots \rangle$ indicates the ensemble average.  
(a) The plots of the mean connectivity $\langle \overline{k}_t\rangle$ for $N=200$. 
It converges to a constant irrespective of the initial value, which is evaluated 
as $\langle \overline{k}_\infty\rangle =   (T/4)^{-1}\sum_{t=(3/4)T}^T \langle\overline{k}_t \rangle\simeq 1.4$. 
The  networks at selected times are presented.  (b) The $N$-dependence of the stationary-state 
mean connectivity. $\langle \overline{k}_{\infty}\rangle\simeq 0.53 +0.17 \ln N$ (solid line) fits 
reasonably the model results (circles). 
The mean connectivity $\overline{k}=L/N$ of the  transcriptional regulatory networks of  four 
species (triangles)~\cite{balazsi05, galanvasquez11, balazsi08,sanz11,balaji06}
and of the bipartite metabolic networks of 506 species (crosses)~\cite{Karp01012005,pkim14} are shown. 
The fitting line (dotted) given by $\langle \overline{k}_{\infty}\rangle\simeq 1.01 +0.15 \ln N$ 
fits the data  of the metabolic networks with $N$ the number of reactions and metabolites.
(c) The cumulative distributions of the in-degree, $C(k)=\langle N^{-1}\sum_{j=1}^N \theta(k_{j}-k)\rangle$ 
at $t=0$ (initial state) and $t=4.8\times 10^5$ (stationary state) for $N=200$. 
The distribution in the random networks of $N=200$ nodes and 
$\langle L\rangle=\langle \overline{k}_\infty\rangle N=1.4N$ links   is also shown for comparison. }
\label{fig:evol_structure}
\end{center}
\end{figure}

The simulation of the proposed model shows a variety of interesting  features of 
evolving networks. Most of all, we find that 
the mean connectivity \REV{$\langle \overline{k}_t\rangle=\langle N^{-1}\sum_{i=1}^N k_i \rangle=\langle L_t\rangle/N$, 
with $k_i = \sum_{j=1}^N A_{ij}$ the in-degree or the number of regulators of node $i$ and }
$L_t$ the total number of links at time $t$, converges to a constant $\langle {\overline{k}}_{\infty}\rangle$, which depends only on 
$N$ regardless of $\overline{k}_0=L_0/N$ [Fig.~\ref{fig:evol_structure} (a)].
The mean connectivity has been shown to converge to $\langle \overline{k}_\infty\rangle=2$  in some 
evolution models~\cite{rohlf07, rohlf08, liu06, PhysRevLett.108.128702,10.1371/journal.pcbi.1002669}, which 
is the critical point distinguishing the ordered and the chaotic phase in random Boolean networks~\cite{derrida86}.
Different values of  $\langle \overline{k}_\infty\rangle$ have been reported in 
other  models~\cite{Sevim2008323,1367-2630-11-3-033005}, where
$\langle \overline{k}_\infty\rangle>2$,  implying 
a fundamental difference between the evolved networks and random networks. 
In our model, $\langle \overline{k}_\infty\rangle$  ranges from $1.2$ to $1.7$ 
for $30\leq N\leq 800$ and 
the data are fitted by a logarithmic growth with $N$ as
$\langle \overline{k}_\infty\rangle\sim 0.53 + 0.17 \ln N$ 
[See  Fig.~\ref{fig:evol_structure} (b)]. 
This suggests that $\langle \overline{k}_\infty\rangle$ would remain small for $N$ reasonably large,
e.g.,  $\langle \overline{k}_\infty\rangle\simeq 2.88$ for $N=10^6$.  
Such sparse connectivity  is identified in real biological 
networks~\cite{balazsi05,galanvasquez11,balazsi08, sanz11,balaji06,Karp01012005,pkim14}. 
The mean connectivities of the transcriptional regulatory networks 
are between $1$ and $3$ while the number of nodes ranges from hundreds to thousands. 
The mean connectivities of the metabolic bipartite networks also range between $1$ and $3$. 
Furthermore, they show logarithmic scaling with $N$ in agreement with our model 
[See Fig.~\ref{fig:evol_structure} (b)].  

The number of regulator nodes \REV{(in-degree)} $k$  is broadly distributed in the evolved network 
compared with the Poissonian distribution of the random networks as seen in Fig.~\ref{fig:evol_structure} (c). 
Such broad distributions are universally observed in  real-world 
networks~\cite{albert02,balazsi05,thieffry98, balaji06,tilee02}. 
The cumulative \REV{in-}degree distribution $C(k)=N^{-1} \sum_{i=1}^N \theta(k_i-k)$, with $\theta(x)$ 
the Heavisde step function, appear to take the form of an exponential function, 
which is   in agreement with the transcriptional regulatory networks of 
{\it S. cerevisiae}~\cite{balaji06,tilee02}. 
\REV{This is, however, inconsistent with the previous studies on the real metabolic networks~\cite{Jeong:2000kx}
or other model networks evolving via node duplication and divergence~\cite{10.1371/journal.pcbi.1002669}, 
which display power-law degree distributions. It is known that the node duplication~\cite{ Vazquez:2003vn} or 
the preferential attachment of links~\cite{Barabasi:1999ys} may lead to such power-law 
degree distributions, which is missing in our model. 
In Ref.~\cite{doi:10.1137/070710111},  the functional form of 
the degree distributions of some real metabolic networks are hard to point out. 

In contrast to the broad in-degree distributions, 
the out-degree $k^{\rm (out)}_i$ in the evolved networks of our model  is found to follow the Poisson distribution as in 
the random networks.  
It is known  that the out-degree distribution is irrelevant to the 
determination of the dynamical phase - ordered or chaotic -  of random Boolean networks~\cite{lee08jpa}.
}

\begin{figure}
\begin{center}
\includegraphics[width=\columnwidth]{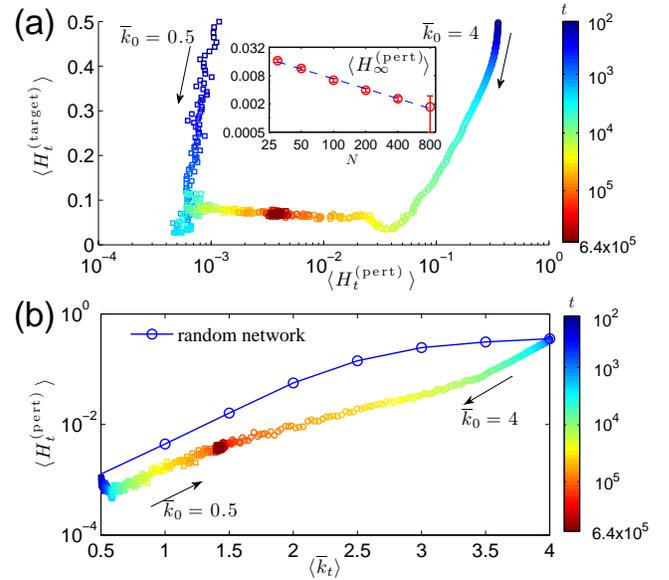}
\caption{(Color online) Time evolution of adaptability and stability. 
(a) Plot of $(\langle H^{\rm (pert)}_t\rangle,\langle H^{\rm (target)}_t\rangle)$ for  
$10^2\leq t<6.4\times 10^5$ and $N=200$ with the initial mean connectivity $\overline{k}_0=4$ 
and $\overline{k}_0=0.5$. \REV{$\mathcal{N}=1000$ simulations are run. }
The color varies with the evolution time $t$ and the arrows indicate the direction of increasing time. 
(Inset) The scaling behavior of the stationary-state Hamming distance $(\langle H^{\rm (pert)}_\infty\rangle$ 
with respect to the number of nodes $N$. $\langle H^{\rm (pert)}_\infty\rangle\sim N^{-0.7}$ (dashed line) 
fits the data. (b) Plots of  $\langle H^{\rm (pert)}_t\rangle$ versus the mean connectivity 
$\langle \overline{k}_t\rangle$ for the evolving networks and the random networks of $N=200$.  
}
\label{fig:evol_dynamics}
\end{center}
\end{figure}

As evolution proceeds, it is more facilitated for the evolving network to get close to or reach 
a given target state. 
Such adaptability is quickly acquired, as implied in the rapid decrease of $\langle H^{\rm (target)}_t\rangle$ 
with increasing $t$ [Fig.~\ref{fig:evol_dynamics} (a)]. 
\REV{We remark that $H^{\rm (target)}_t$ may increase with $t$ even in a single realization of evolution, since the target state, the state demanded by  the environment, may change with time.  }
The extent of perturbation spread $\langle H^{\rm (pert)}_t\rangle$ also decreases rapidly by evolution. 
Its stationary-state value $\langle H^{\rm (pert)}_\infty\rangle$ shows the following 
scaling behavior with $N$:
\begin{equation}
\left\langle H^{\rm (pert)}_\infty\right\rangle\sim N^{-\theta^{\rm (pert)}}, \ \ \theta^{\rm (pert)}\simeq 0.7.
\label{eq:Hpertscal}
\end{equation}
This implies an intermediate level of stability of the evolved networks compared with the following 
networks. 
The random Boolean networks with the mean connectivity at the threshold $\overline{k}_c=2$ find 
the perturbation spread scale similarly to Eq.~(\ref{eq:Hpertscal}) but with a smaller scaling exponent 
ranging between $1/2$ and $1/3$, depending on 
the functional form of the in-degree distribution~\cite{lee08jpa}. 
Therefore, the perturbation spread in those critical random networks is much larger than 
that in the evolved networks for large $N$. 
Figure~\ref{fig:evol_dynamics} (b) shows that  
during the whole period of evolution, the evolving networks 
have smaller spread of perturbation than the random networks with 
the same mean connectivity $\langle \overline{k}\rangle$.    
On the other hand, in a variant of our model, the ``stability-only" model, 
in which only the stability of the wild-type and the mutant is 
evaluated for selection, 
the perturbation spread  scales as $\langle H^{\rm (pert)}_\infty\rangle \sim N^{-1}$ [Fig.~\ref{fig:relax} (b)]. 
The original networks allow larger spread of perturbation than the stability-only model 
in order to facilitate  adaptation to a fluctuating environment. 

The mean connectivity $\langle \overline{k}_\infty\rangle$  is also subject to such a balance constraint. 
As the opposite to the stability-only model, we can consider the ``adaptation-only" model 
in which  only the adaptability of the wild-type and the mutant is considered.
We found that the mean connectivity is much larger  than  in the original 
model.~\footnote{We found that the mean connectivity does not even become 
stationary but  keeps increasing with time in some cases.}
A large number of links make more and larger attractors in the state space, 
which can be helpful for adaptation. 
In the stability-only model,  on the contrary, we find that the mean connectivity is much 
smaller than that of the original model [Fig.~\ref{fig:relax} (a)], suppressing the transitions between attractors. 
All these characteristics demonstrate that the structure and dynamics of the evolved 
networks are  at the boundary between the stable and robust phase and the  flexible and adaptable 
phase~\cite{kauffman69}.

\section{A generalized model}
\label{sec:generalized}

\begin{figure}
\includegraphics[width=\columnwidth]{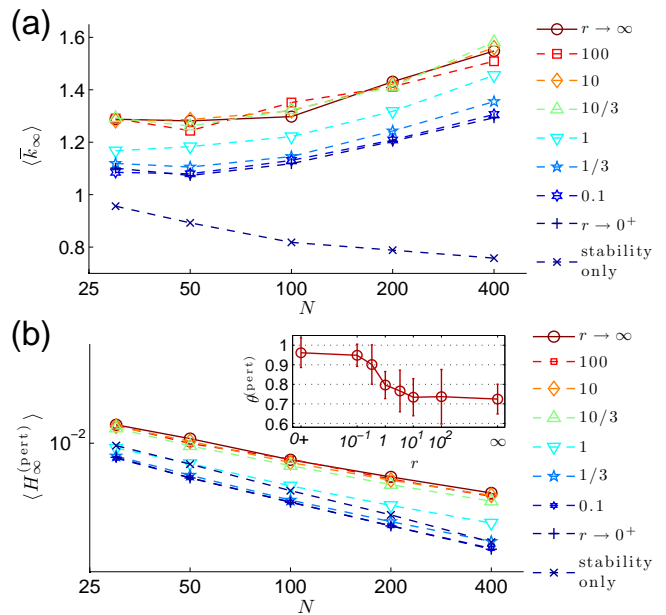}
\caption{(Color online) Mean connectivity and stability  in the generalized model. 
The parameter $r$ is related to the relative importance of adaptability with respect to stability as  in Eq.~(\ref{eq:rdef}). 
 (a) Plots of the stationary-state mean connectivity $\langle \overline{k}_\infty\rangle$ versus the system size $N$.  
$\langle \overline{k}_\infty\rangle$  increases slowly with $N$ for all $r>0$ except for the stability-only model. 
(b)   Plots of the perturbation spread $\langle H^{\rm (pert)}_\infty\rangle$ versus $N$. 
The scaling behavior $\langle H^{\rm (pert)}_\infty\rangle \sim N^{-\theta^{\rm (pert)}}$ is 
observed for all the considered cases. (Inset)  
the scaling exponent $\theta^{\rm (pert)}$  decreases from $1$ to $0.7$ with increasing $r$. 
}
\label{fig:relax}
\end{figure}

In this section, we represent our model in the Hamiltonian approach, which 
offers a natural extension of the model allowing us to check the robustness 
of the obtained results.

The evolution trajectory of the model network corresponds to a path in the space of networks $G$. 
A system of $N$ nodes changes its location in the $G$ space in the stochastic way
as described in Sec.~\ref{sec:model}.
Therefore, a generalized evolution model can be introduced  by specifying
the transition probability $\omega_{G\to G';\Sigma}$ from $G$ to $G'$ 
for a given dynamical state $\Sigma$~\cite{Wagner19961008,Sevim2008323,peixoto12}. 
Note that the dynamical state 
evolves with microscopic time $\tau$ in a deterministic way as long as the network structure $G$ is fixed.  
Suppose that the transition probabilities   satisfy the relation
\begin{align}
{\omega_{G\to G';\Sigma}\over 
\omega_{G'\to G;\Sigma}} = 
\exp & \left(-{H_{G'}^{\rm (target)}-H_{G}^{\rm (target)}\over T^{\rm (target)}}  \right.\nonumber\\
 &\left.  -{{H_{G'}^{\rm (pert)} - H_G^{\rm (pert)} \over T^{\rm (pert)}}}\right),
\label{eq:boltzmann}
\end{align}
where the Hamming distances are computed by Eqs.~(\ref{eq:Htarget}) and (\ref{eq:Hpert}) with 
$\Sigma(t\tau_m) = \Sigma$ and  two temperatures  $T^{\rm (target)}$ and $T^{\rm (pert)}$ are introduced. 
Transitions   to the networks with smaller $H^{\rm (target)}$ and $H^{\rm (pert)}$ are preferred to the extent 
depending on the two temperatures. Our model corresponds to the limit  
\begin{equation}
T^{\rm (target)}\to 0, \ T^{\rm (pert)}\to 0, \ {\rm and}  \  r \equiv {T^{\rm (pert)}\over T^{\rm (target)}}\to \infty, 
\label{eq:rdef}
\end{equation} 
since the transition from $G$ to $G'$ is made only if $H^{\rm (target)}_{G'}< H^{\rm (target)}_G$ 
or $H^{\rm (target)}_{G'}=H^{\rm (target)}_G$ and $H^{\rm (pert)}_{G'}\leq H^{\rm (pert)}_G$. 
In case $T^{\rm (target)}>0$ and $T^{\rm (pert)}>0$, the transition to a less adaptable ($H^{\rm (target)}$-larger) 
or less stable ($H^{\rm (pert)}$-larger) network 
can be made with non-zero probability contrary to that of our model. 
The adaptability-only model corresponds to the limit $T^{\rm (target)}\to 0$ and $T^{\rm (pert)}\to \infty$ 
and the stability-only model  to $T^{\rm (target)}\to \infty$ and $T^{\rm (pert)}\to 0$.

With the transition probabilities satisfying Eq.~(\ref{eq:boltzmann}), each network $G$ appears 
in the stationary state with probability
\begin{equation}
P_{G;\Sigma} \propto \exp\left(- {H_G^{\rm (target)}\over T^{\rm (target)}} -{H_G^{\rm (pert)}\over T^{\rm (pert)}}  \right),
\end{equation}
with the  two Hamming distances playing the role of Hamiltonians coupled with two temperatures. 

To investigate the robustness of the results obtained in Sec.~\ref{sec:evolution}, 
we investigate this generalized model with the temperature ratio  $r$ positive, 
$T^{\rm (pert)}\to 0$, and $T^{\rm (target)}\to 0$.  
For $r>0$, the transition from $G$ to $G'$ is available if and only if 
$H^{\rm (pert)}_{G'}+ r H^{\rm (target)}_{G'}\leq H^{\rm (pert)}_G + r H^{\rm (target)}_G$. 
$r$ controls the relative importance of $H^{\rm (target)}$ with respect to $H^{\rm (pert)}$. 
Simulations show that 
$\langle \overline{k}_\infty\rangle$ displays similar $N$-dependent behaviors for all $r>0$; 
it increases with $N$ slowly [See. Fig.~\ref{fig:relax}(a)]. 
On the contrary, in the stability-only model, 
the mean connectivity decreases with $N$. This highlights the crucial role of adaptation in shaping the 
architecture of regulatory networks. 
Secondly, as shown in Fig.~\ref{fig:relax} (b), 
$\langle H^{\rm (pert)}_\infty\rangle\sim N^{-\theta^{\rm (pert)}}$ with $\theta^{\rm (pert)} \simeq 0.7$ 
is observed not only for $r\to\infty$ but also for sufficiently large $r$, in the range $r\gtrsim 10$. 
For small $r$, roughly $r\lesssim 0.1$ and in the stability-only model, 
$\langle H^{\rm (pert)}_\infty\rangle\sim N^{-1}$, implying that stronger stability 
is achieved than for large $r$. 
The scaling exponent $\theta^{\rm (pert)}$  decreases from $1$ to $0.7$ with  $r$ increasing 
in the range $0.1\lesssim r \lesssim 10$. 
Such robustness of the structural and functional properties  for all large $r$ 
makes our model ($r\to\infty$) appropriate for modeling the evolutionary selection requesting 
both adaptability and stability.

\section{Scaling of fluctuation} 
\label{sec:fluctuation}

As the initial randomly-wired networks evolve,
many of their properties change with time, the investigation of which may illuminate
the mechanisms of evolution by which living organisms optimize their architecture for 
acquiring adaptability and stability.

\begin{figure*}
\includegraphics[width=2\columnwidth]{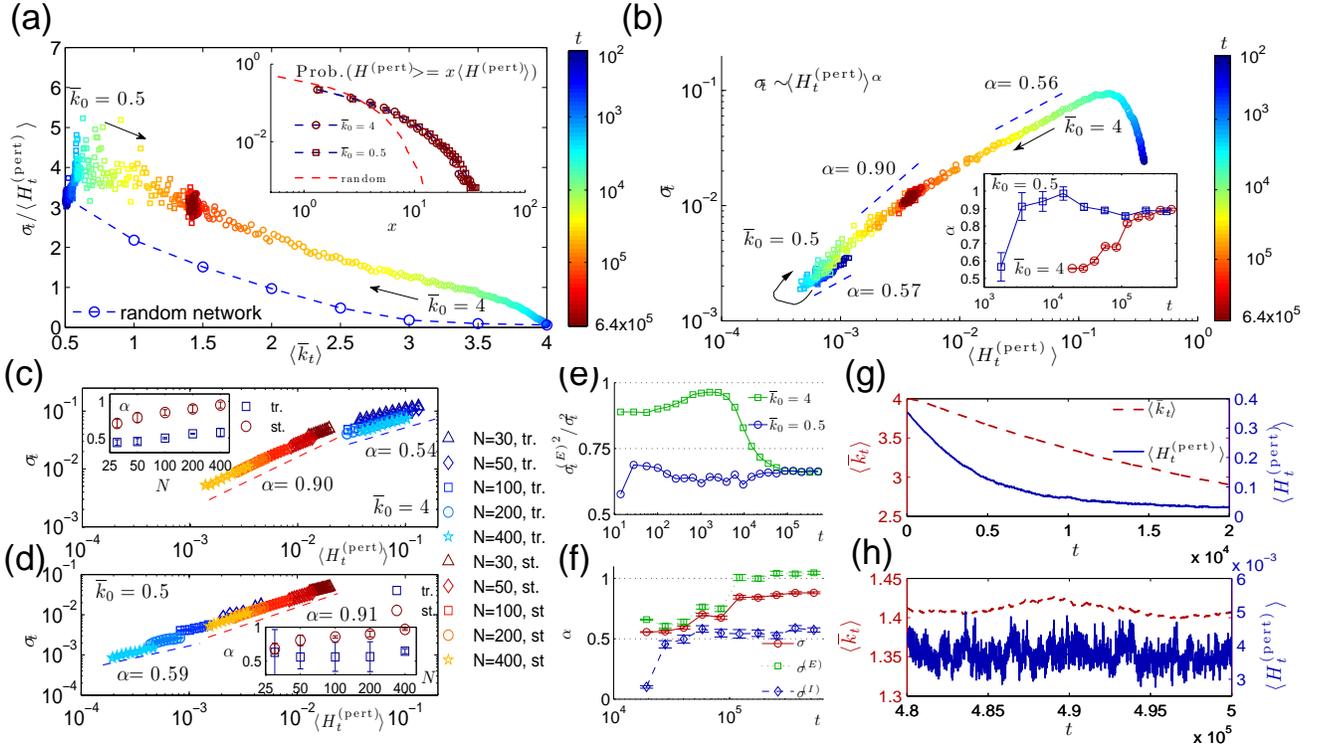}
\caption{(Color online) Scaling behaviors of fluctuation of perturbation spread.  
(a) The normalized fluctuation $\sigma_t/\langle H^{\rm (pert)}_t\rangle$  
as a function of the mean connectivity $\langle \overline{k}_t\rangle$  for $N=200$. 
It  is larger than that in the random networks (dashed line).  
The color varies with the evolution time $t$ and the arrows indicate the direction of increasing time. 
(Inset) The cumulative distributions of $H^{\rm (pert)}_{G,t}$ in the stationary state ($t>4.8\times 10^5$) 
compared with those of the random networks of $\langle \bar{k}\rangle=1.4$ (dashed line). 
(b) Plots of $\sigma_t$ with respect to $\langle H^{\rm (pert)}_t\rangle$ for $\overline{k}_0=4$ and 
$0.5$ and $N=200$.  (Inset) The estimated scaling exponents $\alpha$ in Eq.~(\ref{eq:fluctscal}) 
as functions of time $t$. 
(c) Plots of $\sigma_t$ versus the log-binned values of $\langle H^{\rm (pert)}_t\rangle$ 
in the transient (tr.) and stationary (st.) periods for system sizes $N=50, 100, 200$, and $400$ with  
$\overline{k}_0=4$.  The slopes of the two fitting lines, $0.90$ and $0.54$, are the averages of the 
estimated exponents $\alpha$  in the transient and the stationary period, respectively. 
(Inset)  Plots of $\alpha$ versus $N$ in the transient and the stationary period.  
(d) The same plots as (c) with $\overline{k}_0=0.5$. The slopes, $0.91$ and $0.59$, 
of the two fitting lines are the average of $\alpha$ for  $N=50, 100, 200$, and $400$. 
(Inset) Plots of $\alpha$ versus $N$ in the transient and the stationary periods.
(e) The ratio ${\sigma^{(E)}_t}^2/{\sigma_t}^2$ as  functions of time $t$ for $\overline{k}_0=4$ 
and $0.5$. In the stationary state, ${\sigma_t^{(E)}}^2/{\sigma_t}^2\simeq 0.67$ without 
regards to the initial mean connectivity or the system size.
(f) The estimated scaling exponents $\alpha$  for the whole, external, and internal fluctuation 
at each time $t$ for $N=200$. 
(g) Plots of $\langle H^{\rm (pert)}_t\rangle$ and $\langle \overline{k}_t\rangle$ versus time $t$ 
in the transient period $0<t<20,000$. Both decrease  with little fluctuation. 
(h) Plots of $\langle H^{\rm (pert)}_t\rangle$ and $\langle \overline{k}_t\rangle$ versus $t$ 
in the stationary period $480,000<t<500,000$. 
The  larger fluctuation of $\langle H^{\rm (pert)}_t\rangle$ than $\langle \overline{k}_t\rangle$ is seen. }
\label{fig:fluctuation}
\end{figure*}

Evolution is accompanied by fluctuations.
Environments are different for different groups of organisms and vary with time as well 
even for a given group. 
Mutants are generated at random and thus the specific pathway of evolution 
becomes stochastic. 
The studied networks  also display fluctuations over different realizations of evolution 
$\sigma = \sqrt{\langle A^2\rangle - \langle A\rangle^2}$ for each quantity $A$. 
Among others, here we investigate such ensemble fluctuation  of 
perturbation spread characterizing the system's stability 
$\sigmapert_t  = \sqrt{\left\langle (H^{\rm (pert)}_t)^2\right\rangle - 
\left\langle H^{\rm (pert)}_t\right\rangle^2}$. 
While the evolutionary pressure results in enhancing stability (reducing 
$\langle H^{\rm (pert)}\rangle$),  its fluctuation, normalized by the mean 
$\langle H^{\rm (pert)}\rangle$, is stronger  and 
the whole distribution is broader, respectively, than
those of random networks as shown in Fig.~\ref{fig:fluctuation} (a). 
Such enhancement of fluctuations helps  the evolving network search for the optimal topology 
under fluctuating environments~\cite{eldar10,kussel05,thattai04,wolf05}.  

It is observed for a wide range of real-world systems that  the standard deviation 
$\sigma$ and the mean $m$ of a dynamic variable show the scaling relation 
$\sigma \sim m^\alpha$ with the scaling exponent $\alpha$ reflecting the nature of the dynamical processes: 
For instance, $\alpha=1/2$ in the case of no correlations among 
the relevant variables and their distributions having finite moments 
as in the conventional random walk while the widely varying external influence 
may make such significant correlations as leading to 
$\alpha\neq 1/2$~\cite{menezes04a,menezes04b,meloni08,eisler08}. 
Such scaling relation has been observed for the gene expression level or 
the protein concentration that fluctuates over cells and time~\cite{nacher05,bareven06}. 
Also in our model the  mean $\langle H^{\rm (pert)}_t\rangle$ and the fluctuation $\sigma_t$ of perturbation spread 
at different times $t$  satisfy the scaling relation   
\begin{equation}
\sigmapert_t \sim \left\langle H^{\rm (pert)}_t\right\rangle^\alpha.
\label{eq:fluctscal}
\end{equation}
Interestingly, the scaling exponent $\alpha$ changes with evolution 
[Fig.~\ref{fig:fluctuation} (b)]; 
$\alpha=\alpha_{\rm tr}$ with $\alpha_{\rm tr}\simeq 0.5$ 
for $\bar{k_0}=4$ and 
$\alpha_{\rm tr}\simeq 0.6$ for $\bar{k_0}=0.5$ 
during transient period but  
$\alpha=\alpha_{\rm st}$ with $\alpha_{\rm st}\simeq 1$ 
in the stationary state.
Such crossover in $\alpha$ is robustly observed for all $N$ and $L_0$  
 as shown in Fig.~\ref{fig:fluctuation}(c) and \ref{fig:fluctuation}(d). 

What is the origin of such dynamic crossover in $\alpha$? 
It has been shown that the interplay of exogenous and endogenous 
dynamics may affect the scaling exponent $\alpha$  in systems under the influence 
of external environments~\cite{elowitz02,swain02,menezes04a,menezes04b,meloni08,eisler08}.  
In our evolution model,  the extent of perturbation spread depends on 
the initial perturbation and on the network structure. 
The network structure is the outcome of the specific evolution pathway 
affected by the changing environment.
The location of initial perturbation is determined on a random basis in our model, 
modeling the stochasticity of the internal microscopic dynamics in  real systems. 
Therefore the perturbation spread can be considered as 
 a function of the internal dynamics component $D$ and the network structure $S$, i.e., 
 $H^{\rm (pert)}(D,S)$.
Then the  fluctuation of $H^{\rm (pert)}$ is represented as 
 $\sigma^2 = \langle \langle {H^{\rm (pert)}}^2 \rangle_D \rangle_S -  
{\langle \langle H^{\rm (pert)} \rangle_D \rangle_S}^2$, where 
$\langle \cdots \rangle_D$ and $\langle \cdots \rangle_S$ represent 
the average over $D$ and $S$ as $\int dD P(D)\cdots$ and $\int dS P(S)
\cdots$ and decomposed into  the 
 internal and the external fluctuation as~\cite{elowitz02,swain02}:
\begin{align}
\sigma^2 &= {\sigma^{(I)}}^2 + {\sigma^{(E)}}^2,\nonumber\\
{\sigma^{(I)}}&=\sqrt{\langle \langle {H^{\rm (pert)}}^2\rangle_D\rangle_S 
- \langle {\langle H^{\rm (pert)}\rangle_D}^2\rangle_S},\nonumber\\
{\sigma^{(S)}}&=\sqrt{\langle {\langle H^{\rm (pert)}\rangle_D}^2\rangle_S 
- {\langle \langle H^{\rm (pert)} \rangle_D \rangle_S}^2}.
\label{eq:sigmas}
\end{align} 
The internal fluctuation $\sigmaI$ denotes
the structural average of the internal-dynamics fluctuation of $H^{\rm (pert)}$. 
On the other hand, the external fluctuation $\sigmaE$ is  the structural fluctuation
of the internal-dynamics average  of $H^{\rm (pert)}$. 
In simulations, the quantities $\langle \langle \cdots \rangle_D\rangle_S$ are obtained simply by 
the ensemble averages $\langle \cdots \rangle$. To obtain
$\langle {\langle H^{\rm (pert)}\rangle_D}^2\rangle_S$, 
we use the relation $\langle {\langle H^{\rm (pert)}\rangle_D}^2\rangle_S = 
\langle H^{\rm (pert,I)} H^{\rm (pert,II)}\rangle$~\cite{elowitz02,swain02}, where 
$H^{\rm (pert,I)}$ and $H^{\rm (pert,II)}$ are 
the perturbation spreads from two different initial perturbations on the same network and 
are computed by Eq.~(\ref{eq:Hpert}) 
with different perturbed states $\Sigma^{\rm (pert,I)}$ 
and $\Sigma^{\rm (pert,II)}$ from two initial perturbations.  
Inserting $\langle {H^{\rm (pert)}}^2\rangle = (1/2)(\langle {H^{\rm (pert,I)}}^2\rangle + \langle {H^{\rm (pert,II)}}^2\rangle)$ 
and $\langle {H^{\rm (pert)}}\rangle = (1/2)(\langle {H^{\rm (pert,I)}}\rangle + \langle {H^{\rm (pert,II)}}\rangle)$ 
in Eq.~(\ref{eq:sigmas}), one finds  that  the internal fluctuation is represented as
$\sigmaI^2 =  (1/2) \langle (H^{\rm (pert,I)} - H^{\rm (pert,II)})^2\rangle$ and 
the external fluctuation   is $\sigmaE^2 = \langle (H^{\rm (pert,I)} -\langle H^{\rm (pert,I)}\rangle)
(H^{\rm (pert,II)} -\langle H^{\rm (pert,II)}\rangle)\rangle$.

The external fluctuation 
$\sigma^{(E)}_{t}$ is found to be much larger than  $\sigma^{(I)}_{t}$ for all $t$ [Figure~\ref{fig:fluctuation} (e)], 
implying 
 the wide variation of the network structure arising from exploiting differentiated pathways of evolution 
 in changing environments. 
Moreover, the external fluctuation displays a similar crossover behavior to $\sigma_t$, that is,  
$\sigma^{(E)}_t\sim \langle H^{\rm (pert)}_t\rangle^{\alpha^{(E)}}$ 
with $\alpha^{(E)}$ 
increasing from $\alpha^{(E)}_{\rm tr}$, a value close to $1/2$, 
in the transient period to a value  
$\alpha^{(E)}_{\rm st}\simeq 1$ in the stationary state 
[Fig.~\ref{fig:fluctuation} (f)]. On the other hand, the internal fluctuation behaves as 
$\sigma^{(I)}_{t}\sim \langle H^{\rm (pert)}_t\rangle^{\alpha^{(I)}}$ 
with $\alpha^{(I)}$ remaining close to $1/2$, like in the diffusion process 
[Fig.~\ref{fig:fluctuation}(f)]. 

Which is dominant of the internal and the external fluctuation  has been investigated for 
various complex systems~\cite{menezes04a,menezes04b,meloni08,eisler08}.
Contrary to the static (nature of) systems of the previous works,  
the evolving networks in our model  display a dynamic crossover in the fluctuation 
scaling while the external fluctuation is always dominant. 
To decipher the mechanism underlying this phenomenon,  
we begin with assuming that in the scaling regime the perturbation spread 
$H^{\rm (pert)}_t$
is small and factorized as 
\begin{equation}
H^{\rm (pert)}_t \simeq   D_t S_t, 
\end{equation}
where $D_t$ and $S_t$ are the components reflecting the dependence 
of perturbation spread on the location of initial perturbation and on the global network structure, respectively. 
$D_t$ and $S_t$ are expected to be independent. 
We  assume that their fluctuations scale as  
$\xi^{(D)}_t= \sqrt{\langle D_t^2\rangle - 
\langle D_t\rangle^2}\sim \langle D_t\rangle^{\beta^{(D)}}$ 
and $\xi^{(S)}_t= \sqrt{\langle S_t^2\rangle - \langle S_t\rangle^2}\sim 
\langle S_t\rangle^{\beta^{(S)}}$ with 
$\beta^{(D)}$ and $\beta^{(S)}$  time-independent constants. 
Then the mean of the perturbation spread should be given by 
\begin{equation}
\langle H^{\rm (pert)}_t\rangle = \langle D_t\rangle\langle S_t\rangle 
\label{eq:Hdecomp0}
\end{equation} 
and   the internal and the external fluctuation in Eq.~(\ref{eq:sigmas}) are represented as
\begin{eqnarray}
{\sigma^{(I)}_t} &=& \sqrt{(\langle D_t^2\rangle-\langle D_t\rangle^2)\langle S_t^2\rangle} 
\sim \langle D_t\rangle^{\beta^{(D)}}\sqrt{\langle S_t^2\rangle}, \nonumber\\
{\sigma^{(E)}_t} &=& \langle D_t\rangle \sqrt{\langle S_t^2\rangle  - 
\langle S_t\rangle^2} \sim \langle D_t \rangle \langle S_t\rangle^{\beta^{(S)}}.
\label{eq:fluctdecomp0}
\end{eqnarray}

Using Eqs.~(\ref{eq:Hdecomp0}) and (\ref{eq:fluctdecomp0}), 
we can analyze the scaling behaviors of fluctuations as follows.  
In the transient period before entering the stationary state, 
the network structure  is transformed significantly, 
making the structural component $\langle S_t\rangle$  
essentially govern the perturbation spread 
in its time-dependent behavior, yielding 
\begin{equation}
\langle H^{\rm (pert)}_t\rangle \sim \langle S_t\rangle, \ 
\sigma^{(I)}_t\sim \sqrt{\langle S_t^2\rangle}, \ \sigma^{(E)}_t \sim \langle S_t\rangle^{\beta^{(S)}}.
\end{equation} 
This is supported by the similarity of the 
temporal patterns  of $\langle H^{\rm (pert)}_t\rangle$ and  
the mean connectivity $\langle \overline{k}_t\rangle$ in Fig.~\ref{fig:fluctuation} (g). 
Therefore one can relate the external fluctuation to the mean of perturbation spread as
\begin{equation}
\sigma^{(E)}_t\sim \langle S_t\rangle^{\beta^{(S)}}\sim 
\langle H^{\rm (pert)}_t\rangle^{\beta^{(S)}}.
\end{equation} 
Comparing this with the simulation results in Fig.~\ref{fig:fluctuation}(f), 
we find that $\beta^{(S)}\simeq \alpha^{(E)}_{\rm tr}\simeq 1/2$. 
That is, $\xi^{(S)}\sim \langle S\rangle^{1/2}$. 
The estimated value $\beta^{(S)}$ is also consistent with the simulation result 
$\alpha_{\rm tr}^{(I)}\simeq 1/2$, since  $\sigma^{(I)}_t\sim \sqrt{\langle S_t^2\rangle}\sim 
\sqrt{\langle S_t\rangle^2+{\rm (const.)} \langle S_t\rangle^{2 \beta^{(S)}}}\sim \langle S_t\rangle^{\beta^{(S)}}$, 
with  $\langle S_t\rangle\ll 1$ given $\langle H^{\rm (pert)}\rangle$ small in the scaling regime.

In the stationary state,  the network structure varies little with time;
$\langle \overline{k}_t\rangle$  rarely varies  (Fig.~\ref{fig:fluctuation} (h)).  
In contrast,  $\langle H^{\rm (pert)}_t\rangle$ fluctuates significantly on short time scales. 
This suggests that  randomly-selected locations of initial perturbation, having no 
correlations at different time steps, drive 
such time-dependent behaviors of $\langle H^{\rm (pert)}_t\rangle$. 
Therefore, from 
Eqs.~(\ref{eq:Hdecomp0}) and (\ref{eq:fluctdecomp0}),  the mean and the fluctuation 
of perturbation spread are represented as 
 \begin{equation}
\langle H^{\rm (pert)}_t\rangle \sim \langle D_t\rangle, \ 
\sigma^{(I)}_t\sim \langle D_t\rangle^{\beta^{(D)}}, \ \sigma^{(E)}_t \sim \langle D_t\rangle.
\end{equation} 
Regardless of the value of $\beta^{(D)}$, the external fluctuation is 
proportional to  $\langle H^{\rm (pert)}_t\rangle$, 
\begin{equation}
\sigma^{(E)}_t \sim \langle D_t\rangle \sim \langle H^{\rm (pert)}_t\rangle
\end{equation}
in agreement with the observation $\alpha^{(E)}_{\rm st} \simeq 1$ in 
Fig.~\ref{fig:fluctuation} (f). 
The internal fluctuation is expected to scale as 
$\sigma^{(I)}_t\sim \langle D_t\rangle^{\beta^{(D)}}\sim 
\langle H^{\rm (pert)}_t\rangle^{\beta^{(D)}}$,
which allows us to find
$\beta^{(D)}\simeq\alpha^{(I)}\simeq 1/2$. 
Therefore $\xi^{(D)}\sim \langle D\rangle^{1/2}$ like $\xi^{(S)}\sim \langle S\rangle^{1/2}$.

The above arguments following Eqs.~(\ref{eq:Hdecomp0}) and (\ref{eq:fluctdecomp0}) 
with $\beta^{(S)}\simeq\beta^{(D)}\simeq 1/2$ illustrate 
why the internal fluctuation always scale as $\sigma_t^{(I)}\sim \langle H^{\rm (pert)}_t\rangle^{1/2}$ while
the external  fluctuation shows the dynamic crossover 
from $\sigma_t^{(E)}\sim \langle H^{\rm (pert)}_t\rangle^{1/2}$ to 
$\sigma_t^{(E)}\sim \langle H^{\rm (pert)}_t\rangle$. 
Combined with the observation that the external fluctuation makes a dominant contribution 
to $\sigma_t$, the arguments explain the crossover in the fluctuation scaling of 
perturbation spread shown in Fig.~\ref{fig:fluctuation} (b).

Our results can be compared with the other cases showing a crossover in 
the fluctuation scaling driven by the change of the dominant fluctuation between 
$\sigma^{(I)}$ and $\sigma^{(E)}$~\cite{meloni08}.
On the other hand, $\sigma^{(E)}$ is always dominant in our model. 
The time-varying perturbation spread is dominantly governed by 
the structure component $S_t$  in the transient period and the internal dynamics  
component $D_t$ in the stationary state, 
which underlies the crossover of $\alpha$ from $1/2$ to $1$ in our model. 
The rapid and significant changes of the structure of the evolving networks are identified only in 
the transient period, and the internal stochasticity dominates the statistics of stability in the 
stationary state of evolution. 
Therefore the nature of fluctuations is fundamentally different between the evolved networks 
and the random network or those which are not sufficiently evolved. 

\section{Correlation volume}
\label{sec:correlation}

The evolved networks in our model are more stable than random networks but less stable 
than the stability-only networks as shown by the scaling behaviors of 
$\langle H^{\rm (pert)}_\infty\rangle$ in Sec.~\ref{sec:evolution}. 
Such balance between robustness and flexibility 
is hardly acquired unless the relevant dynamical variables, the spread of perturbation in our 
case, at different sites are correlated with one another.

For a quantitative analysis, 
let us consider the local perturbation $h_{i,t}$ at node $i$ and time $t$  defined as 
\begin{equation}
h_{i,t} = {1\over \tau_m - \tau_s} \sum_{\tau=t\tau_m +\tau_s}^{(t+1)\tau_m}
\left[1-\delta_{b_i(\tau),b_i^{\rm (pert)}(\tau)}\right],
\end{equation}
denoting whether the activity of node $i$ is different between the original state $\Sigma$
and the perturbed state $\Sigma^{\rm (pert)}$. 
Notice that the stability Hamming distance $H^{\rm (pert)}_t$ in Eq.~(\ref{eq:Hpert})
is the spatial average of the local perturbations, $H_t^{\rm (pert)} = 
N^{-1}\sum_{i=1}^N h_{i,t}$.
If node $j$ tends to have larger perturbation than its average when node $i$ does, $h_{i,t}>\langle h_{i,t}\rangle$, their 
local perturbations can be considered as correlated, meaning that local fluctuations at $i (j)$ are
likely to spread to node $j (i)$. 
In that case, we can expect that 
$\langle (h_{i,t} - \langle h_{i,t}\rangle)(h_{j,t} - \langle h_{j,t}\rangle)\rangle= \langle h_{i,t}h_{j,t}\rangle 
-\langle h_{i,t}\rangle\langle h_{j,t}\rangle>0$.  
Therefore we define the correlation volume as
\begin{equation}
\mathcal{C}_{t} \equiv {\sum_{i=1}^N \sum_{j\ne i} 
\left(\langle h_{i,t} h_{j,t} \rangle -\langle h_{i,t} \rangle \langle h_{j,t}\rangle\right)\over
\sum_{j=1}^N \left(\langle h_{j,t}^2\rangle - \langle h_{j,t}\rangle^2\right)}, 
\label{eq:C}
\end{equation}
which represents how many nodes are correlated with a node in the 
perturbation-spreading dynamics. For instance, $\mathcal{C}_t=N-1$ 
if $h_{i,t} = h_{j,t} $ for all $i$ and $j$ (perfect correlation)
and  $\mathcal{C}_t=0$  if the $h$'s are completely independent of one another 
such that $\langle h_{i,t}h_{j,t}\rangle=\langle h_{i,t}\rangle \langle h_{j,t}\rangle$.

One can find that  the variance of the perturbation spread 
$\sigma_t^2 = 
\langle {H^{\rm (pert)}_t}^2\rangle - \langle H^{\rm (pert)}_t\rangle^2$ 
is decomposed into the local variance $\mathcal{S}_t$ and the correlation volume 
$\mathcal{C}_t$ as 
\begin{equation}
\sigma^2_t = \mathcal{S}_t ( 1 +\mathcal{C}_t),
\label{eq:variance}
\end{equation}
where   $\mathcal{S}_t$ is defined in terms of the variance of $h_{i,t}$ as  
\begin{equation}
\mathcal{S}_t \equiv {1\over N^2} 
\sum_{i=1}^N \left(\langle h_{i,t}^2\rangle - \langle h_{i,t}\rangle^2\right).
\end{equation}
The decomposition in Eq.~(\ref{eq:variance}) allows us to see that the fluctuation of perturbation spread 
depends on the magnitude of local fluctuations, $\mathcal{S}_t$, and 
how far the local fluctuation propagates to the system, characterized by 
the correlation volume $\mathcal{C}_t$ in Eq.~(\ref{eq:C}).  
If the $h_{i,t}$'s are independent, the local fluctuation
does not spread, as $\mathcal{C}_t=0$, and the whole variance $\sigma_t^2$ is identical to 
the local variance $\sigma_t^2 = \mathcal{S}_t$. On the contrary, if
the $h_{i,t}$'s are perfectly correlated, the correlation volume is 
$N-1$ and the whole variance $\sigma_t^2$ is $N$ times 
larger than the local variance as 
$\sigma_t^2 = N\mathcal{S}_t$, representing that local fluctuations spread to 
the whole system.

\begin{figure}
\includegraphics[width=\columnwidth]{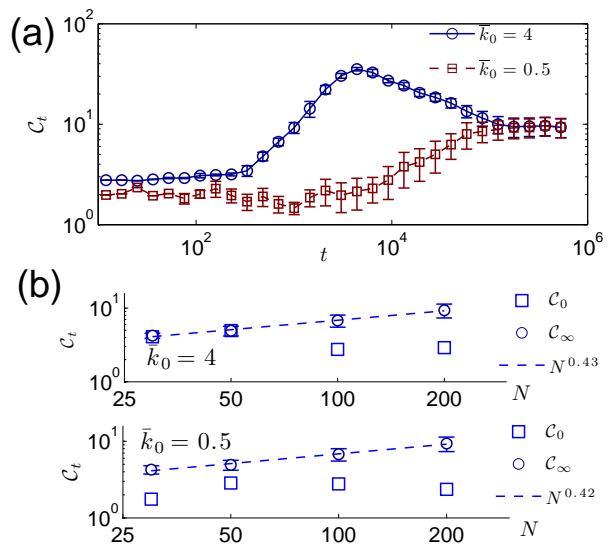}
\caption{(Color online) Correlation volume $\mathcal{C}_t$. 
(a) Plots of $\mathcal{C}_t$ versus time $t$ for $N=200$. 
(b) The initial correlation volume $\mathcal{C}_0$ at $t=0$ and the stationary-state one 
$\mathcal{C}_\infty$ averaged over the stationary period ($t>4.8\times 10^5$) are plotted 
as  functions of the system size $N$ for $\overline{k}_0=4$ (upper) and for $\overline{k}=0.5$ (lower). 
$\mathcal{C}_\infty$ scales with $N$ as $\mathcal{C}_\infty \sim N^{0.43}$ or $N^{0.42}$ 
while $\mathcal{C}_0$ does not increase with $N$.  
}
\label{fig:corrvol}
\end{figure}

In Fig.~\ref{fig:corrvol} (a),  the correlation volume is 
 shown to be larger in the stationary state than in the initial state. 
The correlation volume averaged over the 
stationary period, $\mathcal{C}_\infty$, is about $10$ while that in the initial state, 
$\mathcal{C}_0$, ranges between $2$ and $3$  for $N=200$.  
The dependence of $\mathcal{C}_t$ on the system size $N$ is different 
between the initial and the stationary states. 
Furthermore, the correlation volume in the stationary state  increases with $N$ as 
\begin{equation}
\mathcal{C}_\infty \sim N^\zeta \ {\rm with} \ \zeta \simeq 0.4
\label{eq:cscaling}
\end{equation}
while the correlation volume of the initial network $\mathcal{C}_0$ 
does not increase with $N$ [Fig.~\ref{fig:corrvol} (b)].  
Such a scaling behavior is not seen in the whole fluctuation  $\sigma_t^2$ 
even in the evolved networks.  Therefore, the scaling behavior of the correlation volume 
in Eq.~(\ref{eq:cscaling}) can be another hallmark of the evolved systems and 
can be related to the system's capacity to be stable and adaptable simultaneously. 

\section{Summary and Discussion}
\label{sec:discussion}
In this work we have introduced and extensively investigated the characteristic properties of 
an adaptive network model capturing the generic features of biological evolution.
In reality, the evolutionary selections are made for a population of heterogeneous living organisms, 
as adopted by the genetic algorithm, but here we considered a simplified model, 
where a single network, representing the network structure typical of  a population of organisms, 
add or remove a link depending on whether that change  improves its fitness or not. 
The fitness of a network is evaluated in terms of its adaptability to a changing environment 
and the stability against perturbations in the dynamical state, which look contradictory to each other 
but essential for every living organism.
Despite  such simplification, the model network  reproduces many of the universal network 
characteristics of 
evolving organisms, including the sparsity and scaling of the mean connectivity, broad degree distributions, 
and stability stronger than the random Boolean networks but weaker than the networks evolved 
towards stability only, implying the simultaneous support of  adaptability and robustness. 

Fluctuations and correlations  display characteristic scaling behaviors
in the stationary state of evolution contrasted to those in the transient period or in the initial random-network state. 
The evolutionary pressure drives the regulatory networks towards becoming highly stable 
 by exploiting  different pathways from realization to realization in the rugged fitness landscape, 
 which results in a large fluctuation. 
 The presence of two distinct components in the perturbation-spread dynamics, related to  the different network 
structure depending on the evolution pathway 
and the location of random initial perturbation, respectively,  is shown to bring
the dynamic crossover in the fluctuation scaling. Such evolution makes large correlations as well.

The proposed model is simple and generic allowing us to understand 
the evolutionary origin of the universal features of diverse biological networks. 
It illuminates the nature of dynamic fluctuations and correlations in evolving networks that are 
continuously influenced by the changing environments.  The ensemble of those 
evolving networks can be formulated by the Hamiltonian approach, which depends 
on a time-varying external environment, and thus it opens a way to study biological 
evolution from the viewpoint of statistical mechanics. 
Given the increasing importance of the capacity to manipulate biological systems, 
natural or synthetic, our understanding of biological fluctuations can be particularly useful.  
The strong interaction with environments, like the natural selection in evolution, is 
common to diverse complex systems and thus the theoretical framework to deal 
with multiple components of dynamics presented here  can be of potential use in 
substantiating the theory of complex systems.

\acknowledgments
This work was supported by the National Research Foundation of Korea 
(NRF) grant funded by the Korean Government (MSIP) (No. 2013R1A2A2A01068845).

\end{document}